\PassOptionsToPackage{unicode}{hyperref}
\PassOptionsToPackage{hyphens}{url}
\documentclass[10pt, twocolumn]{article}
\usepackage[english]{babel}

\usepackage{amsmath,amssymb}
\usepackage{iftex}
\usepackage[table]{xcolor} 
\usepackage{colortbl}      
\usepackage{array}         
\usepackage{makecell}
\usepackage{kantlipsum,cuted}

\ifPDFTeX
  \usepackage[T1]{fontenc}
  \usepackage[utf8]{inputenc}
  \usepackage{textcomp} 
\else 
  \usepackage{unicode-math} 
  \defaultfontfeatures{Scale=MatchLowercase}
  \defaultfontfeatures[\rmfamily]{Ligatures=TeX,Scale=1}
\fi
\usepackage{lmodern}
\ifPDFTeX\else
\fi
\IfFileExists{upquote.sty}{\usepackage{upquote}}{}
\IfFileExists{microtype.sty}{
  \usepackage[]{microtype}
  \UseMicrotypeSet[protrusion]{basicmath} 
}{}
\makeatletter
\@ifundefined{KOMAClassName}{
  \IfFileExists{parskip.sty}{%
    \usepackage{parskip}
  }{
    \setlength{\parindent}{0pt}
    \setlength{\parskip}{6pt plus 2pt minus 1pt}}
}{
  \KOMAoptions{parskip=half}}
\makeatother
\usepackage{xcolor}
\usepackage{longtable,booktabs,array}
\usepackage{multirow}
\usepackage{calc} 
\usepackage{etoolbox}
\makeatletter
\patchcmd\longtable{\par}{\if@noskipsec\mbox{}\fi\par}{}{}
\makeatother
\IfFileExists{footnotehyper.sty}{\usepackage{footnotehyper}}{\usepackage{footnote}}
\makesavenoteenv{longtable}
\usepackage{graphicx}
\makeatletter
\newsavebox\pandoc@box
\newcommand*\pandocbounded[1]{
  \sbox\pandoc@box{#1}%
  \Gscale@div\@tempa{\textheight}{\dimexpr\ht\pandoc@box+\dp\pandoc@box\relax}%
  \Gscale@div\@tempb{\linewidth}{\wd\pandoc@box}%
  \ifdim\@tempb\p@<\@tempa\p@\let\@tempa\@tempb\fi
  \ifdim\@tempa\p@<\p@\scalebox{\@tempa}{\usebox\pandoc@box}%
  \else\usebox{\pandoc@box}%
  \fi%
}
\def\fps@figure{htbp}
\makeatother
\ifLuaTeX
  \usepackage{luacolor}
  \usepackage[soul]{lua-ul}
\else
  \usepackage{soul}
\fi
\ifPDFTeX
\fi
\usepackage{bookmark}
\IfFileExists{xurl.sty}{\usepackage{xurl}}{} 
\hypersetup{
  hidelinks,
  pdfcreator={LaTeX}}

\usepackage{authblk}

\title{A theoretical framework for the assessment of water fraction-dependent longitudinal decay rates and magnetisation transfer in membrane lipid phantoms}
\author[1,2]{Heiko Neeb\thanks{neeb@rheinahrcampus.de}}
\author[1]{Felix Schyboll}
\author[3]{Rona Shaharabani}
\author[3]{Aviv A. Mezer}
\author[3]{Oshrat Shtangel}
\affil[1]{Department of Mathematics, Computer Science and Technology, University of Applied Sciences Koblenz, RheinAhrCampus Remagen, Germany}
\affil[2]{Institute for Medical Engineering and Information Processing -- MTI Mittelrhein, University of Koblenz, Germany}
\affil[3]{Edmond and Lily Safra Center for Brain Sciences, Hebrew University of Jerusalem, Israel}

\begin{document}

\twocolumn[
\begin{@twocolumnfalse}

\maketitle

\begin{abstract}
\textbf{Motivation}\\
Phantom  systems consisting of liposome suspensions are widely employed
to investigate quantitative MRI parameters mimicking cellular membranes.
The proper physical understanding of the measurement results, however,
requires proper models for liposomes and their interaction with the
surrounding water molecules. Molecular dynamics (MD) simulations allow
for investigating the basic lipid-water interaction and determining
quantitative MR parameters such as \(R_{1} = 1\text{/}T_{1}\). Here, we
present an MD-based approach for the theoretical prediction of
\(R_{1}\), the dependence of \(R_{1}\) on water concentration and the
magnetization exchange between lipids and interacting water layer in
lipids and lipid mixtures. Moreover, a new parameter is introduced which
quantitatively measures the amount of hydration water based on
conventional spoiled gradient echo MR acquisitions.

\textbf{Methods}\\
Molecular dynamics simulations were performed to determine the native
\(R_{1}\) rates in three lipids and their interacting water pools.
Employing a three-pool exchange model between lipid, hydration water and
free water, the hydration water fraction, \(f_{HW}\), as a new parameter
as well as the magnetization transfer rate between hydration water and
lipids, \(k_{HW,L}\), were quantitatively determined, from which the
water concentration dependence of \(R_{1}\) was predicted for all
liposome systems investigated.

\textbf{Results}\\
Both \(f_{HW}\) and \(k_{HW,L}\) were determined quantitatively from
spoiled gradient echo data by taking the MD-determined relaxation rates
into account. Liposome systems behaved similarly, apart from PLPC which
showed both lower hydration water fraction and lower exchange rate. The
extracted parameters accurately predicted the measured water
fraction-dependent \(R_{1}\) rates.

\textbf{Conclusion}\\
Hydration water fraction and magnetisation transfer between lipids and
water can be determined by a combination of spoiled gradient echo
acquisitions and MD-derived relaxation rates. The parameters enable a
theoretical understanding of MR parameters in liposomes of different
composition.
\end{abstract}

\end{@twocolumnfalse}
]

\newpage

\begin{flushleft}
Abbreviations:

\small{
MRI -- Magnetic resonance imaging

MD -- Molecular dynamics

MR -- Magnetic resonance

PLPC -- 1-Palmitoyl-2-linoleoyl-sn-glycero-3-phosphocholine (16:0/18:2
PC)

NMR -- Nuclear magnetic resonance

DOPS - 1,2-dioleoyl-sn-glycero-3-phospho-L-serine (sodium salt;
18:1/18:1 PS)

MT -- Magnetisation transfer

HW -- Hydration water

FW -- Free water

L -- Lipid

Spg -- Sphingomyelin

Chol -- Cholesterol

MLV -- Multilamellar vesicles

SUV -- Small unilamellar vesicles

MTR -- Magnetisation transfer ratio

VFA -- Variable flip angle

qMT -- Quantitative magnetisation transfer
}
\end{flushleft}

\normalsize

\textbf{\ul{\hfill\break
}}

\section{Introduction}

A thorough understanding of the structural and dynamic properties of
lipid bilayers, as well as transport mechanisms across cell membranes,
is highly significant for biological, pharmaceutical, and diagnostic
purposes \textsuperscript{1}. For example, cell membranes of
oligodendrocytes consisting of lipid bilayers form the myelin in the
central nervous system, the presence of which is of paramount importance
for the transmission of nerve signals \textsuperscript{2}. Changes in
the structure, functioning or lipid composition can lead to the partial
or complete destruction of myelin and are regarded as one of the main
pathophysiological mechanisms in the development of multiple sclerosis
\textsuperscript{3--5}.\\
Various methods are employed for the experimental study of lipid
bilayers, including neutron and photon diffraction measurements,
fluorescence and NMR spectroscopy \textsuperscript{6--9}. Such
techniques allow for the analysis of the dynamics, composition or phase
transitions of bilayer membranes at the nanoscopic scale. For diagnostic
questions, especially in the context of MRI, however, the focus is on
the resulting effects on macroscopically measurable parameters such as
relaxation rates, diffusion of water molecules or the exchange of
magnetisation between lipids and water \textsuperscript{10}. Because
both dynamics as well as structure and composition of the biological
bilayer can be altered in various pathological conditions, a precise
physiological understanding of the effects of such alterations on MR
parameters is necessary to allow for a more specific interpretation of
the measured changes \textsuperscript{11}.\\
However, the situation \emph{in vivo} is usually very complex and
different factors can often equally well account for an observed change
in MRI parameters. For example, a decrease in \(R_{1}\) can be caused by
an increase in free water as well as by a decrease in the magnetisation
exchange rate, e.g. caused by an altered lipid/protein composition
within the plasma membrane. Univariate parameter changes are therefore
usually not very specific, so only the combination of different
quantitative MR parameters allow a more detailed analysis.\\
For a more specific investigation of MRI parameters as well as their
multivariate correlation, specifically prepared \emph{in vitro} phantom
systems are ideal. Shtangel and Mezer have recently proposed a system
that is particularly suitable for the flexible investigation of the
concentration dependence of relaxation rates, diffusion and
magnetisation exchange in bilayer membranes with variable lipid
compositions \textsuperscript{12}. Here, significant differences between
various membrane lipids and lipid compositions such as DOPS and PLPC
were demonstrated for various quantitative MR metrics in that work.\\
Although a phantom system allows for a flexible acquisition of MR
parameters, the interpretation of the measured data is not possible
without a realistic model of the lipid-\(H_{2}O\) interaction and its
effects on the MR observables. Molecular dynamics (MD) simulations,
which allow for a nanoscopic simulation of the proton dynamics both in
bilayer models of lipids and in the interacting water molecules, are
particularly useful for this purpose \textsuperscript{13--15}. Since the
proton dynamics is one of the main determinants of MR relaxation
properties, MD simulations of the lipid-\(H_{2}O\) interaction allow for
a prediction of parameters such as the \(R_{1}\) rate, the diffusion
coefficient or the dipolar contribution to magnetisation transfer.\\
The aim of this work was therefore to adapt the theoretical procedure
described in \textsuperscript{15} to allow for a quantitative
description and interpretation of the \emph{in vitro} measurements of
\textsuperscript{12}. In this way, the experimental data of different
lipids can be quantitatively explained by lipid-specific differences in
proton dynamics, the amount of water bound to the double membrane, and
the exchange of magnetisation between water and lipids. The proposed
model thus complements the experimental approach and allows for a
biophysical interpretation of the measured data as well as a
quantitative determination of fundamental parameters such as the
concentration of hydration water and the magnetisation exchange rate
between lipids and the interacting water pool.

\section{Methods}

\subsection{\normalsize Phantom preparation and MR measurements} 

A phantom system consisting of different liposomes was prepared based on
the hydration-dehydration dry film technique using phospholipids. The
following highly purified and lyophilized lipids were used across
different experiments:
1-Palmitoyl-2-linoleoyl-sn-glycero-3-phosphatidylcholine (PLPC),
cholesterol (all from Sigma-Aldrich), as well as
1,2-dioleoyl-sn-glecero-3-phosphatidylserine 18:1/18:1 (DOPS) and E
sphingomyelin (Spg) (both from LIPOID). Moreover, samples consisting of
3:1 mixtures by weight of PLPC:Chol and 1:1 mixtures of PS-Spg were
created. All samples were diluted with PBS and filled into individual
cuvettes to achieve liposome concentrations between 5\% and 30\% by
volume. The cuvettes were secured to a polystyrene box, which was then
filled with \textasciitilde1\% SeaKem agarose (Ornat Biochemical,
Rehovot, Israel) and \textasciitilde0.0005 M Dotarem (gadoterate
meglumine, Guerbet, Paris, France) in double distilled water to
minimized air-cuvette interfaces for the subsequent MRI measurements.
All details of the lipid phantom system and its experimental setup are
described in \textsuperscript{12}.\\
The different phantom samples were then scanned using a Skyra 3 T MRI
scanner (Siemens, Erlangen, Germany) with a 32-channel receive head coil
at a constant room temperature of \textasciitilde18°C. Specifically,
four 3D spoiled gradient echo (SPGR) datasets with different flip angles
\(\alpha = \left\lbrack 4^{\circ}{,8}^{\circ}{,16}^{\circ}{,30}^{\circ} \right\rbrack\)
were acquired with the following parameters: TE = 3.91 ms, TR = 18 ms,
field of view (FoV) of 205 mm\textsuperscript{2}, and voxel size of 0.5
mm x 0.5 mm x 0.6 mm. T\textsubscript{1} and effective proton density,
\({PD}_{eff}\), were calculated from the acquired variable flip angle
(VFA) data employing the standard Ernst signal model for the SPGR
acquisition and a correction factor for \(B_{1}^{+}\) inhomogeneity.
Finally, the total water concentration, \(c_{H_{2}O}\), was estimated
from the ratio of \({PD}_{eff}\) in the individual samples and in pure
water after correcting for receiver inhomogeneities as detailed
in\textsuperscript{12}.

\subsection{\normalsize Magnetisation transfer model} 

In conventional MRI measurements, information about the properties of
the macromolecular pool can only be extracted indirectly via its
interaction with MR-measurable water molecules. To relate the measured
magnetisation of \textsuperscript{1}H-protons in the \(H_{2}O\) pool to
the magnetisation in biopolymers, the magnetisation transfer dynamics
(MT) between lipids and water can be described as a first-order exchange
process \textsuperscript{16,17}. In the current work, a three-pool model
was thus employed to describe the transfer of magnetisation between
lipids and water as a two-step process: first, MT from lipids to the
adjacent hydration water pool and second, mixing between hydration water
and free water (see Fig. 1). This takes into account the different time
scales, the different transfer mechanisms (chemical exchange and dipolar
coupling vs. diffusion) and the different intrinsic properties of
hydration and free water. The time dependence of the
\(M_{z}\)-components of each pool were thus described by a set of
coupled Bloch-equations:

\begin{strip}
\rule{5cm}{0.2pt}
\begin{eqnarray}
\frac{dM_{FW}}{dt} & = & R_{1,FW}\left( M_{0,FW} - M_{FW} \right) + k_{HW,FW}M_{HW} - k_{FW,HW}M_{FW} \\
\frac{dM_{HW}}{dt} & = & R_{1,HW}\left( M_{0,HW} - M_{HW} \right) - \left( k_{HW,FW} + k_{HW,L} \right)M_{HW} + k_{FW,HW}M_{FW} + k_{L,HW}M_{L} \\
\frac{dM_{L}}{dt}  & = & R_{1,L}\left( M_{0,L} - M_{L} \right) + k_{HW,L}M_{HW} - k_{L,HW}M_{L}
\end{eqnarray}
\hfill \rule{5cm}{0.2pt}
\end{strip}

Here, \(R_{1,i}\) (i=HW, L, FW) denotes the native \(R_{1}\) rate of
hydration water, lipid and free water, respectively, \(k_{L,HW}\) is the
magnetisation exchange rate between lipid and hydration water and
\(k_{FW,HW}\) is the corresponding exchange rate between free water and
hydration water. The inverse rates are given by the principle of
detailed balance, \(M_{0,HW}k_{HW,L} = M_{0,L}k_{L,HW}\), which also
defines the exchange rate normalised to the total pool size,
\(k_{t} = M_{0,HW}k_{HW,L}\). \(M_{0,i}\) characterises the respective
equilibrium magnetisations of the three pools, which sum up to
\(M_{0}\). The hydration water fraction correlates with the lipid
concentration and is defined as the ratio between the equilibrium
magnetisations of the hydration water pool and the lipid pool,
\(f_{HW} = M_{0,HW}\text{/}M_{0,L}\). Here, the equilibrium
magnetisation of the lipid pool,
\(M_{0,L} = M_{0}\left( 1 - c_{H_{2}O} \right)\), is determined by the
total water concentration, \(c_{H_{2}O}\) (see also section A in the
supporting information).

\subsection{\normalsize Molecular dynamics and \(R_{1}\) relaxation rates} 

The determination of intrinsic relaxation rates of lipids and hydration
water is based on the method described in \textsuperscript{15} with
minor extensions to increase the numerical accuracy. For the sake of
completeness, the main steps are described again here. The starting
point is an MD simulation of the proton dynamics of the lipids DOPS,
PLPC and SPG as well as their interacting water pools based on GROMACS,
version 2018.2 \textsuperscript{18,19}. Specifically, lipid double
layers consisting of 100 molecules surrounded by 5000 \(H_{2}O\)
molecules were studied, and their dynamics were simulated at T=291.15K
and normal pressure by the CHARMM36 all-atom force field
\textsuperscript{20} using the TIP4p-FB model as described in
\textsuperscript{15}. Proton trajectories at \(5 \cdot 10^{5}\)
different time points were determined with a time interval of
\(\Delta t = 2ps\), resulting in a total simulation time of \(1\mu s\).\\
The trajectories of the \textsuperscript{1}H nuclei formed the starting
point for the determination of \(R_{1}\) rates. \(R_{1}\) can be
calculated in first-order perturbation theory from the basic equations
of quantum mechanics, leaving as the only unknown the spatio-temporal
dynamics of the protons, which can be expressed by correlation functions
\textsuperscript{21}. Specifically, the correlation functions of first-
and second-order spherical harmonics weighted by the inverse cubic
distance are required. Using the proton trajectories from the MD
simulation, these functions were first determined for a selected
\textsuperscript{1}H nucleus. For this purpose, all pairs between the
selected proton and each of the other protons in the respective pool
(lipid or water) were formed. For each pair, a separate correlation
function was determined for both first- and second-order spherical
harmonics, which were then summed up for the selected proton. Finally,
the proton-specific functions were averaged over all the protons in the
pool. A possible offset due to non-averaged residual dipolar couplings
was subtracted and then Fourier-transformed at \(\omega_{0}\)
(first-order) and at \(2\omega_{0}\) (second-order). The resulting
spectral densities are the only unknowns in the analytical solutions for
\(R_{1}\), so \(R_{1}\) was calculated separately for each lipid as well
as its associated water pool \textsuperscript{15}. Finally, \(R_{1}\) of
the 1:1 lipid mixture PS-SPM was determined by the average of the two
lipid-specific \(R_{1}\)s, while \(R_{1}\) for PC-Chol was approximated
by the relaxation rate of PLPC, since cholesterol represents only 25\%
of the total weight fraction.\\
The size of the water pool, consisting of 50 molecules per lipid
molecule, was deliberately chosen large enough to allow for an accurate
determination of the correlation functions of rapidly tumbling
\textsuperscript{1}H nuclei. For a rapid exchange between free and
hydration water, the MD determined decay rate of the full water pool,
\(R_{1,H2O}^{eff}\), is thus given by (see also Fig. S1 in the
supporting information),

\begin{equation}
	R_{1,H2O}^{eff} = c_{HW}R_{1,HW} + \left( 1 - c_{HW} \right)R_{1,FW}
\end{equation}

Based on \(R_{1,H2O}^{eff}\), \(R_{1,HW}\) was determined by assuming
\(R_{1,FW} = 0.3Hz\) for the free water pool as described below. The
relative fraction of hydration water in Equation (4), \(c_{HW}\), was
extracted from the simulated proton density distribution of lipid and
water pools (Fig.). Specifically, the boundary between free and
hydration water was defined as the distance at which the proton density
had decreased to less than 95\% of its value in free water. The
resulting hydration water fractions (\(c_{HW}\)) of all lipids are
listed in Table 1.

\begin{figure}[t!]
\centering
\begingroup
\includegraphics[width=2.6in]{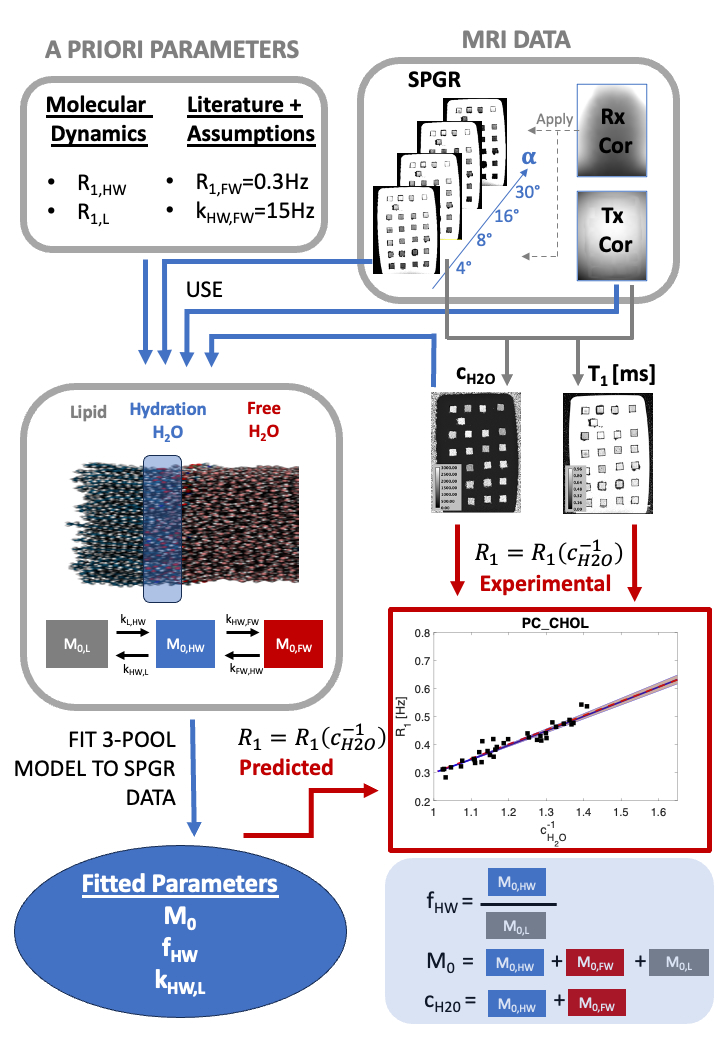}
\caption{\textit{Workflow employed for the determination of hydration water fraction, $f_{HW}$, magnetisation exchange rate between hydration water and lipid, $K_{HW,L}$, as well as the $R_1$ relaxivity of different lipids, $R_1(c_{H_2O}^{-1})$. Spoiled gradient echo data acquired at four different flip angles were fitted to a three-pool exchange model. The parameters space of the model was constrained by the prior definition of pool specific $R_1$ rates, employing both molecular dynamics simulations as well as literature values. Based on the fitted parameters, the functional dependence $R_1(c_{H_2O}^{-1})$ was theoretically determined and compared to experimentally measured data on $R_1$ and water concentration, $c_{H_2O}$. The latter were obtained by the approach described in \textsuperscript{12}.}}
\label{fig:methods_overview}
\endgroup

\end{figure}

\subsection{\normalsize Determination of model parameters} 

Equations 1-3 contain a total of 14 free parameters. The relaxation
rates of hydration water and lipids were extracted from the molecular
dynamics simulation as detailed above, while the longitudinal relaxation
rate of free water was set to \(R_{1,FW} = 0.3Hz\). This value was
determined from the extrapolation of the relaxation data to
\(c_{H_{2}O} = 1\) (see Results, i.e. Fig. 3) and is consistent with the
\(R_{1}\) range of pure water reported in the literature
\textsuperscript{15,22}. Since magnetisation exchange between free water
and hydration water is diffusive and therefore fast on the NMR
timescale, a fixed value of \(k_{FW,HW} = 15Hz\) was assumed in the
following. This is greater than the typical \(R_{1}\) rates of both
water pools, thereby defining an effective fast exchange regime.
Nevertheless, the results do not depend on the specific choice of
\(k_{FW,HW}\) as long as the fast exchange condition is met. Finally,
the total water concentration (\(c_{H_{2}O}\)) was determined by the MRI
measurement as described above, leaving only three unknown parameters
remaining: the total magnetisation, \(M_{0}\), the exchange rate,
\(k_{HW,L}\), and the hydration fraction, \(f_{HW}\) (see section A of
the supporting information).\\
To determine the unknown parameter values, the steady-state
z-magnetisations, \(M_{i}^{SS}\) (i=HW, FW, L), of four spoiled gradient
echo acquisitions at flip angles
\(\alpha = \left\lbrack 4^{\circ}{,8}^{\circ}{,16}^{\circ}{,30}^{\circ} \right\rbrack\)
were calculated by analytical solving Eqs. 1-3 \textsuperscript{23}. The
theoretical MR signal was then given by the sum of the transverse
magnetisations of both water pools, i.e.
\(\left( M_{HW}^{SS} + M_{FW}^{SS} \right)sin\alpha\), whereas the
magnetisation of the lipid pool was neglected due to its short
\(T_{2}^{*}\). The four theoretical signal intensities were fitted to
the corresponding four experimental signal intensities of each
individual data point (cuvette) by employing a restricted least squares
method. Experimental data were normalised to match the normalisation
condition employed in the signal model (section B in the supporting
information). The permissible parameter range was limited to values for
\(k_{HW,L}\) from {[}0 to 36{]} Hz and \(f_{HW}\) from {[}0.03 to
0.66{]}, thereby covering the whole regimes between no exchange and fast
exchange as well as between sparsely and fully hydrated bilayers.
Although the experimental data were normalised to \(M_{0} = 1\),
\(M_{0}\) was left as a free parameter due to the noise of the
experimental data but was restricted to a very narrow range of {[}0.98,
1.02{]}. Cases where the result coincided with one or more of the three
lower or upper interval boundaries were excluded from the analysis. For
the remaining datapoints, the mean and standard errors of the three
parameters (\({\overline{M}}_{0}\), \({\overline{f}}_{HW}\),
\({\overline{k}}_{HW,L}\)) were determined individually for each lipid
or lipid mixture.

\subsection{\normalsize Simulations} 

\begin{figure*}[t!]
\centering
\begingroup
\includegraphics[width=6.6in]{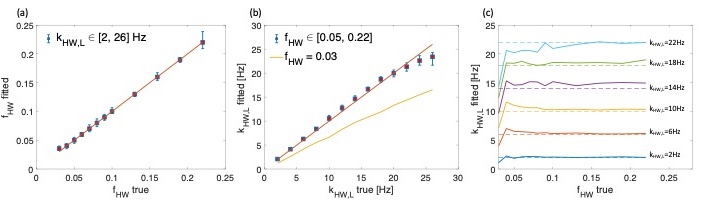}
\caption{\textit{Results from the validation of the three-pool model. (a)
shows the fitted hydration water fraction versus the true value of
\(f_{HW}\) for different exchange rates, \(k_{HW,L}\). The corresponding
results for the exchange rate between hydration water and lipid is shown
in (b). The data were averaged over hydration water fractions ranging in
the interval {[}0.05 0.22{]}. In addition, the orange line shows the
results for a small \(f_{HW} = 0.03\), which clearly deviates from the
results at \(f_{HW} \geq 0.05\). In both (a) and (b), the red line shows
the identity of fitted and true parameter values. To further illustrate
the interdependence between \(f_{HW}\) and \(k_{HW,L}\), (c) shows the
fitted results of the exchange rate as a function of hydration water
fraction for different true values of \(k_{HW,L}\) (colored lines).}}
\label{fig:Result_Validation}
\endgroup
\end{figure*}

To assess the precision and accuracy of the fitted parameters, synthetic
data were generated based on the analytical expression for the
steady-state magnetisation of a spoiled gradient echo sequence using the
MR sequence parameters stated above \textsuperscript{12,23}. For the
simulated datasets, the exchange rate \(k_{HW,L}\) ranged between 2Hz to
24 Hz while \(f_{HW}\) varied between 0.03 to 0.24 (Fig. S2). In all
cases, \(M_{0} = 1\) and a water concentration of \(c_{H2O} = 0.75\)
were assumed. The latter value was chosen as it represents the typical
average water concentration in the experimental samples studied here.
Gaussian noise with a standard deviation of 1/SNR was added, where
SNR=5828 corresponds to the raw signal-to-noise ratio in the
experimental data, i.e. in the \(TR \rightarrow \infty\) limit. The
large SNR value is due to the fact that each measured datapoint
(cuvette) is an average over all associated voxels (on average
\(N_{Voxel} = 1091\)), leading to an increased SNR by a factor
\(\sqrt{N_{Voxel}}\).\\
For each combination of the above parameters, the sum of the transverse
magnetisations of the two water pools was again determined, and the
optimisation procedure described in (D) was performed. The fitted
parameters were then compared to their nominal values.

\subsection{\normalsize Relaxivity and water concentration dependent R\textsubscript{1}} 

The relaxivity of the lipids, \(dR_{1}\text{/}dc_{H_{2}O}^{- 1}\), is
given by the water concentration dependence of the longitudinal
relaxation rate, \(R_{1}\left( c_{H_{2}O} \right)\). To determine the
function \(R_{1}\left( c_{H_{2}O} \right)\) theoretically, further
synthetic data sets were generated following the procedure described in
(E). However, the water concentration was not assumed to be fixed here,
but varied in the interval \(c_{H_{2}O} \in \lbrack 0.4,1\rbrack\). For
each value of \(c_{H_{2}O}\) and for each flip angle, the sum of the
transverse magnetisations of both water pools was calculated from which
\(R_{1}\) was calculated, in complete analogy to the \(R_{1}\)
determination in the experimental data (see also Fig. 1). It is
important to notice that the magnetisation was calculated based on
parameters \(f_{HW}\) and \(k_{HW,L}\) which were different from the
values given in Table 1 (\({\overline{f}}_{HW}\) and
\({\overline{k}}_{HW,L}\)). While \({\overline{f}}_{HW}\) and
\({\overline{k}}_{HW,L}\) were calculated by averaging all available
datapoints, both parameters were redetermined here by employing only
datapoints with total water concentration ranging between
\(c_{H2O} = 0.7\) and \(c_{H_{2}O} = 0.85\). This procedure was chosen
to avoid an overfitting and to investigate the validity of the parameter
values outside the water concentration ranges that were employed for
their determination.\\
Finally, a comparison with the fast proton diffusion (FPD) model
proposed by Fullerton et al. was performed, which also assumes a fast
exchange between hydration water and free water, similar to the
three-pool model employed here \textsuperscript{24}. To determine the
associated model parameters \(C_{1}\) and \(C_{2}\), the function

\begin{equation}
	R_{1}\left( c_{H_{2}O} \right) = C_{1} \cdot \frac{1}{c_{H_{2}O}} + C_{2}
\end{equation}

was fitted to the experimental \(R_{1}\) values at different water
concentrations. In the notation of the current work, the parameters are
given by \(C_{1} = f_{HW}^{FPD}\left( R_{1,HW} - R_{1,FW} \right)\) and
\(C_{2} = R_{1,FW} - C_{1}\). Employing the \(R_{1}\) rates of free and
hydration water (see box in the left top of Fig. 1), the hydration water
fraction, \(f_{HW}^{FPD}\), can thus be determined in the FPD model.
However, it should be noted that \(R_{1,HW}\) in Fullterton's model
represents the effective \(R_{1}\) rate of the hydration water, which
already includes the contribution of the magnetisation exchange with the
lipid pool. In the present work, \(R_{1,HW}\) is the pure \(R_{1}\) rate
of the HW pool and the MT effects were taken into account by explicit
modelling magnetisation exchange through equations 1-3.

\section{Results}

\begin{table*}[t]
\centering

\begingroup

\setlength{\tabcolsep}{4pt} 
\renewcommand{\arraystretch}{2.5} 

\begin{tabular}{l c c c c c c c c c}
    \cline{2-10}
     \rule{0pt}{-20pt} \multirow{2}{*}{} & \multicolumn{4}{|c|}{\cellcolor{lightgray} {\textbf{Molecular Dynamics}}} & \multicolumn{5}{c|}{\cellcolor{gray} {\textbf{Model Fit to Experimental Data}}} \\ 
  
	  \renewcommand{\arraystretch}{4.5}
     
     \rule{0pt}{-20pt} & \cellcolor{lightgray} $R_{1,L}$[Hz] & \cellcolor{lightgray} $R_{1,HW}$[Hz] & \cellcolor{lightgray} $c_{HW}$ & \cellcolor{lightgray} $N_{Layers}$ & \cellcolor{gray} $N_{Curv}$ & \cellcolor{gray} $\overline{f}_{HW}$ & \cellcolor{gray} $\overline{f}_{HW}^{FPD}$ & \cellcolor{gray} $\overline{k}_{HW,L}$[Hz] & \cellcolor{gray} $\overline{k}_{t}$[Hz]\\ \hline
     PLPC & 4,4 & 4,2 & 0,34 & 6,4 & 20 (39) & \makecell{0,069 \\ ±0,005} & \makecell{0,054 \\ ±0,004} & 
     \makecell{6,8 \\ ±1,4} & \makecell{0,069 \\ ±0,01} \\

	 DOPS & 3,9 & 4,3 & 0,33 & 6,1 & 9 (14) & \makecell{0,13 \\ ±0,007} & \makecell{0,14 \\ ±0,006} & 
	 \makecell{14,6 \\ ±1,9} & \makecell{0,46 \\ ±0,04} \\
	 
	 SPG & 3,2 & 4,9 & 0,26 & 5,7 & 12 (13) & \makecell{0,18 \\ ±0,01} & \makecell{0,17 \\ ±0,006} & 
	 \makecell{6,3 \\ ±1,3} & \makecell{0,23 \\ ±0,03} \\
	 
	 PC-Chol & 4,4 & 4,2 & 0,34 & 6,4 & 14 (35) & \makecell{0,11 \\ ±0,008} & \makecell{0,11 \\ ±0,003} & 
	 \makecell{21,6 \\ ±2,7} & \makecell{0,47 \\ ±0,07} \\
	 
	 PS-SPG & 3,55 & 4,6 & 0,3 & 5,9 & 5 (7) & \makecell{0,15 \\ ±0,017 \\ (0,16 \\ ± 0,006)} & 
	 \makecell{0,15 \\ ±0,021 \\ (0,16 \\ ±0,004)} & 
	 \makecell{11,8 \\ ±4,7 \\ (10,5\\ ±1,2)} & 
	 \makecell{0,27 \\ ±0,04 \\(0,34 \\ ±0,03)} \\

	\hline
\end{tabular}

\endgroup

\caption{\textit{Results from the molecular dynamics simulation (left
panel) and from the three-pool model fit to the spoiled gradient echo
data (right panel) for the different lipids and lipid mixtures. The MD
study allows for the determination of the native \(R_{1}\) rate of
lipids (\(R_{1,L}\)), \(R_{1}\) of hydration water (\(R_{1,HW}\)), the
relative amount of hydration water in the water pool simulated
(\(c_{HW}\), see Eq. 4) and the corresponding number of hydration layers
(\(N_{Layers}\)). The latter has been determined from the ratio of the
thickness of the dark blue area in Fig. S1 and the diameter of a single
water molecule \textsuperscript{34}. The experimental results were based
on the numbers of cuvettes that could be fitted (\(N_{Curv}\)), whereas
the total number of measured cuvettes is given in brackets.
\({\overline{f}}_{HW}\) is the average hydration water fraction obtained
by the fit of the three-pool model, \({\overline{f}}_{HW}^{FPD}\) the
corresponding result from the FPD model \textsuperscript{24},
\({\overline{k}}_{HW,L}\) characterizes the average exchange rate
between hydration water and lipid (relative to the hydration water pool
size) and \({\overline{k}}_{t}\) is the corresponding exchange rate
normalized to the total magnetisation of all pools.}}
\label{tab:results}
\end{table*}

\subsection{\normalsize Simulation results} 

The results of the molecular dynamics simulation are shown in Table 1.
The native \(R_{1}\) rates of all studied lipids were in a comparable
range, between 3.2 Hz for SPG and 4.4 Hz for PLPC. Similarly, the
\(R_{1}\) of the hydration water, \(R_{1,HW}\), and the number of
hydration layers, \(N_{Layers}\), showed only minor differences. Results
from the validation of the exchange model are shown in Figure 2. It can
be noticed that the hydration water fraction can be determined
accurately over a large range of \(k_{HW,L}\) (Fig. 2a). In contrast,
the exchange rate fit showed a systematic underestimation for exchange
rates \(k_{HW,L} \geq 18Hz\) (Fig. 2b and 2c). These deviations are more
pronounced at low hydration water fractions, resulting in a large bias
for \(f_{HW} \leq 0.03\) (orange curve in Fig. 2b). Nevertheless, the
hydration water fraction was well above that value for all the lipids
investigated here (Table 1).

\begin{figure*}[t!]
\centering
\begingroup
\includegraphics[width=6.6in]{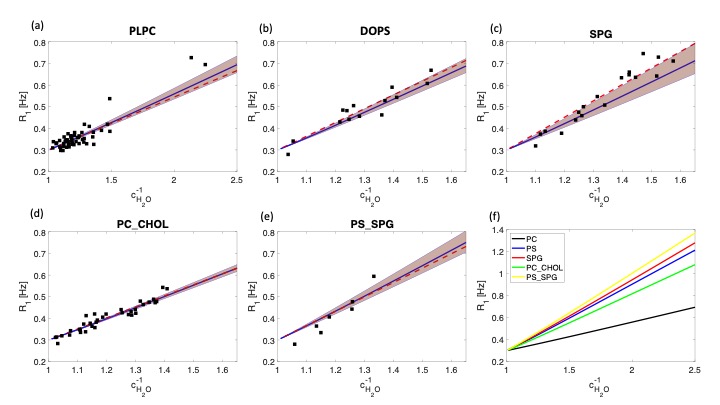}
\caption{\textit{R\textsubscript{1}-relaxivity of different lipids and
lipid mixtures. The dependence of \(R_{1}\) on the inverse water
concentration is shown for (a) PLPC, (b) DOPS, (c) SPG, (d) PC-Chol and
(e) PS-SPG. The filled squares show the experimental results of
individual cuvettes, while the blue curves present the prediction of the
three-pool exchange model. The uncertainty of the predicted \(R_{1}\) is
covered by the filled red area. Note that the uncertainty increases with
decreasing water concentration for all lipids and lipid mixtures
investigated. Moreover, the results from a linear fit to the data are
superimposed as dashed red lines. Their slope defines the hydration
water fraction in the FPD model (see text). For a better and direct
comparison, the three-pool model predictions for all lipids and mixtures
are shown in (f).}}
\label{fig:Result_Relaxivity}
\endgroup
\end{figure*}

\subsection{\normalsize Exchange Model} 

The results of the magnetisation exchange model fit are shown on the
right half side of Table 1. PLPC had both a significantly lower
hydration water fraction and a lower exchange rate as compared to the
other lipids and lipid mixtures. However, the addition of 25\%
cholesterol to PLPC was sufficient to triple its exchange rate.
Moreover, \(f_{HW}\) increased from 0.069 to 0.11, so that after the
addition of cholesterol both parameters showed values comparable to
those of the other lipids. For the second lipid mixture examined,
PS-SPG, fitted values of both \(f_{HW}\) and \(k_{HW,L}\) agreed with
the weighted average of the corresponding parameter values of DOPS and
SPG. Finally, it should be noted that the number of cuvettes that could
be fitted was rather different for the different samples. While for
PC-Chol less than half and for PLPC only about half of all acquired data
points were included, only a single data point had to be removed from
the analysis for SPG. Possible reasons for this are discussed in detail
below.

\subsection{\normalsize $R_1$-Relaxivity, $R_1(c_{H_2O})$} 

Figure 3 shows the water fraction dependence of \(R_{1}\). In all the
lipid systems, a good agreement between the measured values and the
model predictions was observed. Again, it is obvious from Fig. 3 that
PLPC behaved differently from the other lipids and that the addition of
cholesterol resulted in a significant increase in relaxivity. All lipids
showed a linear relationship between \(R_{1}\) and inverse water
fraction, which is also consistent with the prediction of the FPD model.
The linear fit of the data points thus allowed for the determination of
the two parameters \(C_{1}\) and \(C_{2}\) in Eq. 5 (dashed red line in
Fig. 3). The resulting values for the hydration water fraction
(\(f_{HW}^{FPD}\)) were consistent with the three-pool model employed in
this work as shown in Table 1.

\section{Discussion}

\subsection{\normalsize Hydration water fraction} 

Results of the molecular dynamics simulation demonstrated a similar
behaviour of all investigated lipids at the nanoscopic level, both with
respect to their native \(R_{1}\), the hydration water fraction as well
as the \(R_{1}\) rate of the hydration water. However, the fit of the
exchange model showed that the hydration water fraction of PLPC was
significantly lower than that of other lipids. This is in stark contrast
to the MD-based results, which predicted that the number of hydration
layers is largest in PLPC (see Table 1). One has to consider, however,
that only a small part of a bilayer was modelled here. The MD-based
results thus represent the expected behaviour on a nanoscopic level
whereas potential microstructural differences cannot be captured by that
approach. Therefore, the observed difference between molecular dynamics
simulation and the three-pool model fit might point to microstructural
differences between the different lipids that could lead to a reduction
of the hydration water fraction, such as a reduced thickness of the
water layer between two layers of multilamellar vesicles (MLV).
Similarly, a higher proportion of micelles or small unilamellar vesicles
(SUV) and the associated increase in lipid volume relative to the
surface in PLPC might explain the reduced \(f_{HW}\). However, we do not
have experimental evidence for any of these scenarios to have happened.\\
Moreover, systematic errors due to the specific MR acquisition protocol
and the incomplete/improper magnetisation exchange model have to be
considered as potential causes for the small value of \(f_{HW}\) in
PLPC. Specifically, incomplete spoiling of the magnetisation could cause
a systematic bias given the short TR of the SPGR data. This would mostly
affect the \(\alpha = 30^{\circ}\) measurements, where the contribution
of refocusing pathways contribute is strongest \textsuperscript{25,26}.
Since both \(T_{1}\)and \(T_{2}\) (data not shown) were higher in the
PLPC suspensions than in the suspensions of the other lipids and lipid
mixtures, a systematic error due to non-perfect spoiling of stimulated
and spin echoes should be particularly noticeable in PLPC. To
investigate this effect, the complete analysis was repeated without the
\(\alpha = 30^{\circ}\) measurement. Even in such a reduced data set, it
is still possible to measure \(f_{HW}\) with high accuracy (see Fig.
S3). Moreover, the basic observation with the lowest value for
\(f_{HW}\) in PLPC did not change, even though \(f_{HW}\) increased
slightly in all samples (Table S1).\\
The different behaviour of PLPC could also be related to an incomplete
magnetisation exchange model. In fact, Zhang et al. recently published a
\textsuperscript{17}O-NMR study describing for the first time the very
different lifetimes of surface-near \(H_{2}O\) molecules in DMPC
\textsuperscript{6}. Contrary to the previous assumption of a very short
\(H_{2}O\) residence time in the nano-to picosecond range, their study
showed that only water interacting with the outer choline group is in
fast exchange with the free water. On the other hand, MD simulations
tend to significantly underestimate water lifetimes, often by many
orders of magnitude \textsuperscript{27}. This interpretation is
consistent with the behaviour of PC-Chol observed in this study. The
addition of 25\% cholesterol results in a nearly doubling of the surface
water fraction in comparison to PLPC (see Table 1). Cholesterol leads to
the formation of potholes in the bilayer surface with the phosphate
group in a myelin bilayer extending about 5 Ångstroms further into the
free water region than the hydroxyl group of cholesterol
\textsuperscript{28,29}. This would support the assumption that a slow
exchange of water between the phosphate group and the choline group, and
thus the free water pool, is responsible for the low value of
\(f_{HW}\), since the formation of potholes in the area of cholesterol
increases the lipid surface and in consequence the concentration of
hydration water. In addition, the altered surface morphology will likely
contribute to a more rapid direct exchange between the phosphate group
hydration water and the free water. However, it has to be taken into
account that DMPC, in contrast to DOPC, is a saturated lipid with
different acyl chain dynamics, different phase behaviour and altered
mechanical properties of the formed bilayer \textsuperscript{8}. In
particular, the addition of cholesterol is not expected to lead to the
formation of raft domains and the associated change in surface
morphology and surface dehydration in DOPC-Chol bilayers at 20°C
\textsuperscript{9}.\\
In order to properly capture exchange effects on different time scales,
the exchange model needs to be extended to include a slow-exchanging or
non-exchanging pool of water. If the different \(H_{2}O\) lifetimes in
the head region were also to be confirmed for other lipids, the
conclusion of this study would change in that \(f_{HW}\) characterises
only the part of hydration water which is in fast exchange with the
surrounding free water pool. The determination of the total hydration
water concentration would require an extension of the exchange model,
which likely leads to more noise in the fit parameters due to the low
number of experimental data points. Furthermore, it is doubtful whether
such a pool could be resolved experimentally at all, given the very
similar relaxation rates of tightly bound water and lipids.\\
The predictions of the three-pool model were compared to the fast proton
diffusion (FPD) model in this work, giving comparable results for the
hydration water fraction in both cases. Although the two approaches are
based on a fast exchange between hydration- and free water, the FPD
model assumes that no magnetisation is exchanged between hydration water
and biopolymers. As discussed in detail below, the explicit inclusion of
this parameter allows for a quantitative determination of the exchange
rate between lipids and hydration water. This is expected to be
specifically relevant at lower field strength, given the power-law
increase of lipid \(R_{1}\) with decreasing \(B_{0}\), which results in
a more rapid relaxation dynamics as \(B_{0}\) decreases
\textsuperscript{15}. Moreover, studies on supported lipid bilayers have
demonstrated the presence of different lateral diffusion regimes within
a bilayer membrane upon dehydration \textsuperscript{30}, thereby
affecting relaxation times due to the altered proton dynamics. Such
effects can directly be captured by MD simulations performed here. It
should also be noted that the FPD approach requires at least two samples
with different water concentrations to perform the fit, in contrast to
the three-pool model. This could be particularly relevant if the
approach is adopted for the investigation of \emph{in vivo} data in
future studies.

\subsection{\normalsize Magnetisation exchange} 

The measured exchange rates suggest a slow to intermediate magnetisation
transfer between lipid and hydration water and a fast exchange between
HW and FW, in analogy to the FPD model. This observation is consistent
with a recently published \emph{in vivo} study, in which the parameters
of a complex exchange network consisting of different layers of
\(H_{2}O\) and myelin were determined \textsuperscript{16}. The results
described in this study fit well with our results on the
slow/intermediate exchange between lipid and hydration water and the
fast exchange with free water, even if they contradict older
measurements as discussed in detail in \textsuperscript{16}.\\
The exchange rates determined here were consistent with independent
measurements of the magnetisation transfer ratio, MTR
\textsuperscript{12}. While DOPS and PC-Chol show the smallest MTR at
comparable levels, magnetisation exchange in SPG is significantly
reduced. The largest MTR is found in suspensions containing PLPC. This
hierarchy is consistent with the results for \({\overline{k}}_{t}\) in
Table 1. Moreover, the exchange rates align with those presented in
recent quantitative MT SPGR-model experiments conducted with membrane
lipid \textsuperscript{31}. In particular, the addition of cholesterol
significantly increases the magnetisation transfer between a lipid and
its interacting hydration water pool. Such a behaviour has been known
since the early 1990s \textsuperscript{28}. Although the specific
mechanism of the increased MT observed in the early studies was later
questioned, Kucharczyk et al. have confirmed that the addition of
cholesterol to unsaturated POPC MLV results in an approximately doubled
MT at physiological pH, with an even greater effect at acid pH
\textsuperscript{32}. A recent double quantum coherence filtered NMR
study confirmed the result also in saturated DMPC bilayers
\textsuperscript{33}.\\
Moreover, the experimentally determined \(k_{HW,L}\) and \(f_{HW}\) in
PS-SPG were consistent with the values obtained by averaging the
corresponding parameters of the individual lipids (see Table 1, last
row). This suggests that no significant cross-interaction exists between
both lipids, at least none that would cause a change in the hydration
water fraction or the magnetic exchange properties.\\
The absolute values of \(k_{HW,L}\) were smaller when the
\(\alpha = 30^{\circ}\) measurement was removed from the analysis (Table
S1). Furthermore, more data points were fitted without producing
artificial results. The theoretical model can thus better describe the
experimental data, which highlights the possible contribution of
non-perfect spoiling of the residual magnetisation in the large flip
angle measurements (see above). Despite the generally decreasing
exchange rates, the basic hierarchy between the lipids was maintained,
although the effect of cholesterol addition was somewhat reduced. This
could be due to the fact that the replacement of PLPC by cholesterol
leads to a prolongation of the \(T_{2}\) time and thus a greater
contribution of refocusing pathways in cases of non-ideal spoiling. The
lower values of \(k_{HW,L}\) in the reduced data set, on the other hand,
had little effect on the predicted dependence of \(R_{1}\) on water
concentration (see Figure S4). This shows that the relaxivity is
determined mainly by the hydration water fraction and only to a smaller
extent by \(k_{HW,L}\).\\
In general, the results demonstrate that a quantitative determination of
exchange rates and hydration water fraction with variable flip-angle
(VFA) measurements is feasible, provided that sufficiently precise a
priori information on the relevant relaxation rates is available. Such
an approach for the quantitative determination of exchange rates could
be particularly relevant for \emph{in vivo} qMT imaging at high field
strengths since VFA measurements can be performed with low \(\alpha\),
while completely avoiding the application of saturation pulses.
Moreover, relaxation rates of biopolymers decrease with increasing
\(B_{0}\), so that uncertainties have a smaller impact than at lower
field strengths. Further improvements are possible, especially since the
current acquisition protocol was not specifically designed for the
quantitative determination of exchange rates.

\subsection{\normalsize Limitations} 

The main limitations of the current study have already been discussed
above, namely the neglection of hydration water pools with different
residence times within the lipid head groups and the systematic errors
that could result from non-perfect spoiling at large flip angles.\\
Moreover, we did not determine the native \(R_{1}\) rate of cholesterol
which was assumed to equal that of PLPC in the corresponding mixture. It
should be noted that cholesterol represents only 25\% by mass in the
PC-Chol suspensions studied and does not form bilayers, so a
determination of its native \(R_{1}\) was not meaningful in the same way
as for the other lipids. Nevertheless, a molecular dynamics simulation
of the combined PC-Chol bilayer would in principle have been possible
for the determination of an effective \(R_{1}\). This approach is
meaningful as long as the intermolecular spin cross-relaxation is fast
enough to ensure rapid mixing of magnetisation between the two lipid
species. Ideally, this investigation should thus be performed localised
by calculating separate \(R_{1}\) rates for positions of cholesterol and
the positions of the corresponding PLPC molecules. Additionally, an
explicit determination of cross-relaxation rates and the spin diffusion
coefficient should be performed to validate the assumption of a fast
mixing of magnetisation. However, such an approach is computationally
very expensive and was not directly feasible due to hardware
limitations. Therefore, possible systematic errors due to the assumption
of identical native \(R_{1}\) rates in PLPC and PC-Chol cannot be ruled
out completely.

\section{Conclusion}

In the present work, a theoretical model based on the nanoscopic
description of lipid-water interactions was presented to explain
measured quantitative MR data of lipid suspensions. It was shown that
two parameters, the fraction of hydration water, \(f_{HW}\), and the
magnetisation exchange rate between hydration water and lipid,
\(k_{HW,L}\), can be quantitatively determined based on the acquisition
of standard spoiled gradient echo sequences with different flip angles.\\
We have observed that the hydration water fraction in PLPC suspensions
was significantly lower than in the other lipids and that the addition
of cholesterol resulted in a clear increase of \(f_{HW}\). Moreover, the
\(R_{1}\)-relaxivity of different lipids was mainly determined by
\(f_{HW}\) and only to a minor extent by the magnetisation transfer
rate. Thus, the amount of hydration water is likely a very important
factor for \(R_{1}\) \emph{in vivo} as well, which further highlights
the potential for a quantitative determination of \(f_{HW}\).\\
The method presented here was intended to provide a theoretical
explanation for the data measured in \textsuperscript{12}. Therefore,
the acquisition protocol was not optimised for the determination of
\(k_{HW,L}\) and \(f_{HW}\), but rather for the precise and accurate
determination of \(T_{1}\) and proton density. Adaptations of the
protocol are expected to lead to improvements in the precision of
\(f_{HW}\) and \(k_{HW,L}\), which also could facilitate their future
investigation \emph{in vivo}.

\section{Acknowledgements}

The authors thank Jaqueline Groll for her support in sorting the
experimental data so that they can efficiently be employed for model
fitting.

\section{Data availability}

Data and analysis scripts can be obtained from the corresponding author
upon reasonable request.

\section{References}

1. Ammendolia DA, Bement WM, Brumell JH. Plasma membrane integrity:
implications for health and disease. \emph{BMC Biol}. 2021;19(1):71.
doi:10.1186/s12915-021-00972-y

2. Kister A, Kister I. Overview of myelin, major myelin lipids, and
myelin-associated proteins. \emph{Front Chem}. 2023;10:1041961.
doi:10.3389/fchem.2022.1041961

3. Shaharabani R, Ram-On M, Avinery R, et al. Structural Transition in
Myelin Membrane as Initiator of Multiple Sclerosis. \emph{J Am Chem
Soc}. 2016;138(37):12159-12165. doi:10.1021/jacs.6b04826

4. Filippi M, Bar-Or A, Piehl F, et al. Multiple sclerosis. \emph{Nat
Rev Dis Primers}. 2018;4(1):43. doi:10.1038/s41572-018-0041-4

5. Lassmann H. Multiple Sclerosis Pathology. \emph{Cold Spring Harb
Perspect Med}. 2018;8(3):a028936. doi:10.1101/cshperspect.a028936

6. Zhang R, Cross TA, Peng X, Fu R. Surprising Rigidity of Functionally
Important Water Molecules Buried in the Lipid Headgroup Region. \emph{J
Am Chem Soc}. 2022;144(17):7881-7888. doi:10.1021/jacs.2c02145

7. Costigan SC, Booth PJ, Templer RH. Estimations of lipid bilayer
geometry in £uid lamellar phases. \emph{Biochimica et Biophysica Acta}.
Published online 2000.

8. Nanda H, García Sakai V, Khodadadi S, Tyagi MS, Schwalbach EJ, Curtis
JE. Relaxation dynamics of saturated and unsaturated oriented lipid
bilayers. \emph{Soft Matter}. 2018;14(29):6119-6127.
doi:10.1039/C7SM01720K

9. M'Baye G, Mély Y, Duportail G, Klymchenko AS. Liquid Ordered and Gel
Phases of Lipid Bilayers: Fluorescent Probes Reveal Close Fluidity but
Different Hydration. \emph{Biophysical Journal}. 2008;95(3):1217-1225.
doi:10.1529/biophysj.107.127480

10. Neema M, Stankiewicz J, Arora A, Guss ZD, Bakshi R. MRI in Multiple
Sclerosis: What's Inside the Toolbox? \emph{Neurotherapeutics}.
2007;4(4):602-617. doi:10.1016/j.nurt.2007.08.001

11. Weiskopf N, Edwards LJ, Helms G, Mohammadi S, Kirilina E.
Quantitative magnetic resonance imaging of brain anatomy and in vivo
histology. \emph{Nat Rev Phys}. 2021;3(8):570-588.
doi:10.1038/s42254-021-00326-1

12. Shtangel O, Mezer AA. A phantom system for assessing the effects of
membrane lipids on water proton relaxation. \emph{NMR in Biomedicine}.
2020;33(4):e4209. doi:10.1002/nbm.4209

13. Schyboll F, Jaekel U, Petruccione F, Neeb H. Origin of
orientation‐dependent R \textsubscript{1} (=1/T \textsubscript{1} )
relaxation in white matter. \emph{Magnetic Resonance in Med}.
2020;84(5):2713-2723. doi:10.1002/mrm.28277

14. Falck E, Patra M, Karttunen M, Hyvönen MT, Vattulainen I. Lessons of
Slicing Membranes: Interplay of Packing, Free Area, and Lateral
Diffusion in Phospholipid/Cholesterol Bilayers. \emph{Biophysical
Journal}. 2004;87(2):1076-1091. doi:10.1529/biophysj.104.041368

15. Schyboll F, Jaekel U, Petruccione F, Neeb H. Dipolar induced
spin-lattice relaxation in the myelin sheath: A molecular dynamics
study. \emph{Sci Rep}. 2019;9(1):14813. doi:10.1038/s41598-019-51003-4

16. Van Gelderen P, Duyn JH. White matter intercompartmental water
exchange rates determined from detailed modeling of the myelin sheath.
\emph{Magnetic Resonance in Med}. 2019;81(1):628-638.
doi:10.1002/mrm.27398

17. Pike GB. Pulsed magnetization transfer contrast in gradient echo
imaging: A two‐pool analytic description of signal response.
\emph{Magnetic Resonance in Med}. 1996;36(1):95-103.
doi:10.1002/mrm.1910360117

18. Bauer P, Hess B, Lindahl E. GROMACS 2022.4 Manual (2022.4). Zenodo.
Published online 2022. https://doi.org/10.5281/zenodo.7323409

19. Abraham MJ, Murtola T, Schulz R, et al. GROMACS: High performance
molecular simulations through multi-level parallelism from laptops to
supercomputers. \emph{SoftwareX}. 2015;1-2:19-25.
doi:10.1016/j.softx.2015.06.001

20. Vanommeslaeghe K, Hatcher E, Acharya C, et al. CHARMM general force
field: A force field for drug‐like molecules compatible with the CHARMM
all‐atom additive biological force fields. \emph{J Comput Chem}.
2010;31(4):671-690. doi:10.1002/jcc.21367

21. Fischer MWF, Majumdar A, Zuiderweg ERP. Protein NMR relaxation:
theory, applications and outlook. \emph{Progress in Nuclear Magnetic
Resonance Spectroscopy}. 1998;33(3-4):207-272.
doi:10.1016/S0079-6565(98)00023-5

22. Tsukiashi A, Min KS, Kitayama H, et al. Application of
spin-crossover water soluble nanoparticles for use as MRI contrast
agents. \emph{Sci Rep}. 2018;8(1):14911. doi:10.1038/s41598-018-33362-6

23. Spencer RGS, Fishbein KW. Measurement of Spin--Lattice Relaxation
Times and Concentrations in Systems with Chemical Exchange Using the
One-Pulse Sequence: Breakdown of the Ernst Model for Partial Saturation
in Nuclear Magnetic Resonance Spectroscopy. \emph{Journal of Magnetic
Resonance}. 2000;142(1):120-135. doi:10.1006/jmre.1999.1925

24. Fullerton GD, Potter JL, Dornbluth NC. NMR relaxation of protons in
tissues and other macromolecular water solutions. \emph{Magnetic
Resonance Imaging}. 1982;1(4):209-226. doi:10.1016/0730-725X(82)90172-2

25. Hennig J. Echoes---how to generate, recognize, use or avoid them in
MR‐imaging sequences. Part II: Echoes in imaging sequences.
\emph{Concepts Magn Reson}. 1991;3(4):179-192.
doi:10.1002/cmr.1820030402

26. Hennig J. Echoes---how to generate, recognize, use or avoid them in
MR‐imaging sequences. Part I: Fundamental and not so fundamental
properties of spin echoes. \emph{Concepts Magn Reson}.
1991;3(3):125-143. doi:10.1002/cmr.1820030302

27. Paulino J, Yi M, Hung I, et al. Functional stability of water
wire--carbonyl interactions in an ion channel. \emph{Proc Natl Acad Sci
USA}. 2020;117(22):11908-11915. doi:10.1073/pnas.2001083117

28. Koenig SH. Cholesterol of myelin is the determinant of gray‐white
contrast in MRI of brain. \emph{Magnetic Resonance in Med}.
1991;20(2):285-291. doi:10.1002/mrm.1910200210

29. Koenig SH, Brown RD, Spiller M, Lundbom N. Relaxometry of brain: Why
white matter appears bright in MRI. \emph{Magnetic Resonance in Med}.
1990;14(3):482-495. doi:10.1002/mrm.1910140306

30. Chattopadhyay M, Krok E, Orlikowska H, Schwille P, Franquelim HG,
Piatkowski L. Hydration Layer of Only a Few Molecules Controls Lipid
Mobility in Biomimetic Membranes. \emph{J Am Chem Soc}.
2021;143(36):14551-14562. doi:10.1021/jacs.1c04314

31. Shtangel O, Mezer AA. Testing quantitative magnetization transfer
models with membrane lipids. \emph{Magnetic Resonance in Med}. Published
online June 14, 2024:mrm.30192. doi:10.1002/mrm.30192

32. Kucharczyk W, Macdonald PM, Stanisz GJ, Henkelman RM. Relaxivity and
magnetization transfer of white matter lipids at MR imaging: importance
of cerebrosides and pH. \emph{Radiology}. 1994;192(2):521-529.
doi:10.1148/radiology.192.2.8029426

33. Yang W, Lee J, Leninger M, Windschuh J, Traaseth NJ, Jerschow A.
Magnetization transfer in liposome and proteoliposome samples that mimic
the protein and lipid composition of myelin. \emph{NMR in Biomedicine}.
2019;32(7):e4097. doi:10.1002/nbm.4097

34. Schatzberg P. Molecular diameter of water from solubility and
diffusion measurements. \emph{J Phys Chem}. 1967;71(13):4569-4570.
doi:10.1021/j100872a075

\end{document}